\documentclass[a4paper,aps,prl,showpacs,twocolumn,superscriptaddress]{revtex4}
\usepackage{graphicx,t1enc}

\begin{document}

\title{Non-local Interaction Effects on Pattern Formation in Population
Dynamics}
\author{M. A. Fuentes}
\affiliation{Consortium of the Americas for Interdisciplinary Science and
Department of Physics and Astronomy, University of New Mexico, Albuquerque, NM
87131, U.S.A.} \affiliation{Centro At{\'o}mico Bariloche and Instituto
Balseiro, 8400 S. C. de Bariloche, Argentina}
\author{M. N. Kuperman}
\affiliation{Consortium of the Americas for Interdisciplinary Science and
Department of Physics and Astronomy, University of New Mexico, Albuquerque, NM
87131, U.S.A.} \affiliation{Centro At{\'o}mico Bariloche and Instituto
Balseiro, 8400 S. C. de Bariloche, Argentina}
\author{V. M. Kenkre}
\affiliation{Consortium of the Americas for Interdisciplinary Science and
Department of Physics and Astronomy, University of New Mexico, Albuquerque, NM
87131, U.S.A.}

\begin{abstract}
\ \newline \ \newline We consider a model for population dynamics such as for
the evolution of bacterial colonies which is of the Fisher type but where the
competitive interaction among individuals is non-local, and show that spatial
structures with interesting features emerge. These features depend on the
nature of the competitive interaction as well as on its range, specifically on
the presence or absence of tails in, and the central curvature of, the
influence function of the interaction.
\end{abstract}

\maketitle

\vspace{1cm}


Population dynamics covers a wide spectrum of fields. The formation of
patterns in the evolution of bacterial colonies provides an example
\cite{jaco,waki,lin,nels,thesis}. Such studies, e.g., the growth of an initial
nucleus of cells, are important within a clinical framework since the control
of
bacterial infections might be better performed if the dynamics of the growth
of bacteria were understood at a basic level. The Streptococcus pyogenes or
Group A streptococcus, that has been made famous in recent times by media
reports of flesh-eating bacteria which claim the lives of up to 25\% of its
victims, provides one example. Processes related with cancer development
provide another.

When dealing exclusively with the evolution of cell population and its
spatio-temporal features, it is usual to focus  attention on some processes
such as reproduction, competition for resources, and diffusion, and to neglect
others such as mutation. The Fisher equation is, therefore, a good starting
for such studies. We show below new results concerning pattern formation that
arise from a natural extension of the Fisher equation. This differential
equation considers, besides the diffusion process with coefficient $D$, growth
of the population at rate $a$ and a limiting process controlled by $b$,
associated generally with competition or struggle for resources \cite{fish}:
\begin{equation}
\frac{\partial u\left( x,t\right) }{\partial t}=D\frac{\partial ^{2}u\left(
x,t\right) }{\partial x^{2}}+au\left( x,t\right) -bu^{2}\left( x,t\right).
\label{originaleq}
\end{equation}

Our generalization consists in incorporating {\it non-local } effects in the
competition terms:
\begin{eqnarray}  \label{influeq}
\frac{\partial u\left( \vec x,t\right) }{\partial t} & = & D\nabla^2 u\left(
\vec x,t\right)+a\,u(\vec x,t) \\
& & -b\,u(\vec x,t) \int_{\Omega} u(\vec y,t) f_\sigma ( \vec x,  \vec y).
\nonumber
\end{eqnarray}
Here $f_{\sigma}(\vec x,\vec y)$ is a positive function or distribution which
we call the {\it influence function}, characterized by a range $\sigma$, and
normalized in the domain $\Omega$ under investigation. The physical origin of
non-local aspects in the competition interaction is easy to understand. For
instance, in the case of bacteria, the diffusion of nutrients and/or the
release of toxic substances can cause non-locality in the interaction. While
non-local competition has been mentioned earlier \cite{lee,mog}, we show below
consequences of the non-local terms that have not been reported earlier,
specifically, striking features of the dependence of the patterns on the
nature
as well as  range of the influence function.

It is known that no patterns appear in the extreme local limit $f_\sigma(\vec
x,\vec y)=\delta(\vec x-\vec y)$ which reduces (\ref{influeq}) to
(\ref{originaleq}). No patterns appear in the extreme nonlocal limit either,
where $f_\sigma(x)$ is a constant, as can be seen from an explicit analytical
solution given by one of the present authors \cite{vmkpasi}:
\begin{equation}
u( x,t) =\left[ e^{-at} +%
\frac{b \overline{u}_0}{a}\left( 1-e^{-at}\right) \right] ^{-1}\int \Psi (
x-y,t) u_0(y) dy,
\end{equation}
where $\Psi \left( z,t\right) $ is the standard Gaussian propagator $(4\pi
Dt)^{-1/2}exp(-z^{2}/4Dt)$ of the diffusion equation, and $\overline{u}_0$ is
the integral over all space of the initial density $u_0(x)$. In the
intermediate case, however, patterns do appear. The simplest forms for the
influence function are a Gaussian and a square distribution. The latter is
simply a normalized constant function within a range and vanishing outside.
The
former is given by
\begin{equation}
f_{\sigma }(x,y)=\alpha (x) \,\exp \left[-\frac{(x-y)^{2}}{2\sigma
^{2}}\right] \label{ifunction}
\end{equation}%
in the 1-dimensional case. In the case of periodic boundary conditions with
spatial period $L$, the normalization factor $\alpha (x)$  is a constant,
$\frac{1}{\alpha} =\sqrt{\frac{\pi}{2}}\, \sigma \, \mbox{erf}\left
(\frac{L}{\sqrt{2}\sigma}\right)$, whereas, for zero-flux boundary conditions,
it is space-dependent:
\begin{equation}
\frac{1}{\alpha(x)} =\sqrt{\frac{\pi}{2}}\, \sigma \, \left[
   \mbox{erf}\left(\frac{ x}{\sqrt{2}\sigma}\right)
   -\mbox{erf}\left(\frac{ x- L}{\sqrt{2}\sigma}\right) \right].
 \label{cdneumman}
\end{equation}
Is is helpful to characterize the influence function by its width at the
origin. For periodic boundary conditions, in which case $f_\sigma$ depends on
the difference $z=x-y$, this width is given by $\Sigma=\left[
-d^2(\mbox{ln}(f_\sigma)/dz^2)_{z=0}\right]^{-1/2}$, equals $\sigma$ for a
Gaussian, and  is infinite  for the square function.

\begin{figure}[ht]
\centering \resizebox{\columnwidth}{!}{\rotatebox[origin=c]{-90}{
\includegraphics{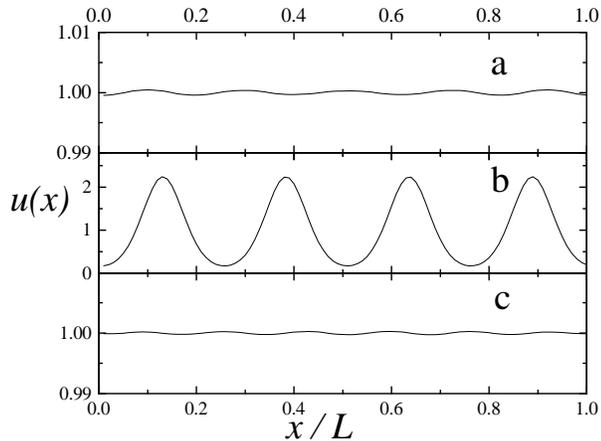}}}
\caption{Dramatic difference in the amplitude of the steady state patterns
with Gaussian influence function and periodic boundary conditions, for
$\Sigma/L=$ a) $0.205$, b)$0.245$ and c)$12$. Here, $u(x)$ is plotted
in units of $a/b$, and $x$ in units of $L$. } \label{gauss1dper}
\end{figure}

\begin{figure}[ht]
\centering \resizebox{\columnwidth}{!} {\rotatebox[origin=c]{-90}{
\includegraphics{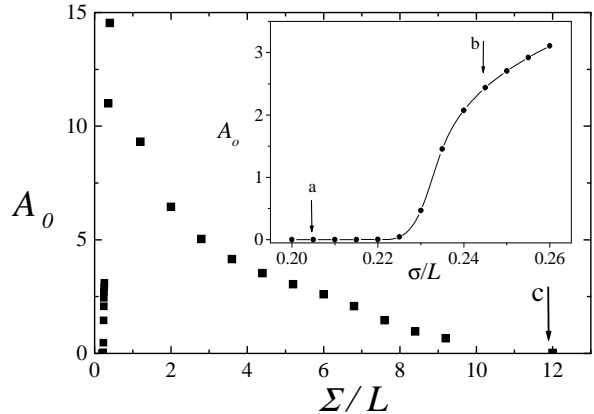}}}
\caption{Dependence of the amplitude of the patterns on the ratio of the width
of the influence function to the domain size. Inset, which depicts the small
$\Sigma/L$ behavior on an expanded scale, shows an apparent transition around
the value $2/9$ of $\Sigma/L$. The arrows mark the three patterns depicted in
Fig. 1.} \label{ampli}
\end{figure}

Our calculations show that, for the case of the square influence
function (infinite $\Sigma$),
patterns of nontrivial amplitude appear for all values of the cut-off interval
of the square, the amplitude of the peaks being not uniform but presenting a
periodic modulation. For a Gaussian influence
function (finite $\Sigma$), the patterns exhibit a curious feature in the
periodic boundary condition
case. Two critical values of the width $\Sigma$ are seen separating trivial
patterns with vanishing amplitude from nontrivial patterns with substantial
amplitude \cite{foot}. This is displayed in Figs. \ref{gauss1dper} and
\ref{ampli}.
The critical width depends linearly on the domain size $L$ (with a slight
deviation for small domains) making it possible to display our results with
$\Sigma/L$ on the x-axis. The results displayed are for a system size $L$ of
100 sites
with $D=1\times 10^{-3}$, $a=1$, $b=1$, in arbitrary but consistent units.
A plot of the pattern amplitude $A_0$
(see Fig. \ref{ampli}) shows a striking transition around the value $2/9$ of
the ratio $\Sigma/L$.

Is this sharp difference between square and Gaussian influence functions the
result of the cut-off inherent in the former? In particular, could the sudden
rise of the
pattern amplitude in the inset of Fig. 2 occur because of the onset of the
natural cut-off imposed on the Gaussian by the finite domain size? In order to
answer these questions, we considered domains large enough (so that the domain
size was unimportant) and used a combination of a Gaussian and a square
influence function, i.e., a Gaussian of width $\Sigma,$ (equivalently, range
$\sigma$), multiplied by a pulse  so that it vanishes abruptly beyond a
cut-off $x_c.$  We found results which suggest that the non-negligible
patterns are indeed associated with the cut-off nature of a square
function or the cut-off imposed on a Gaussian by the finite size of
the domain: large-amplitude patterns appeared for  cut-off  Gaussian
for \emph{any} value of
$\Sigma$ provided $x_c$ was small enough.

\begin{figure}[ht]
\centering \resizebox{\columnwidth}{!} {\rotatebox[origin=c]{-90}{
\includegraphics{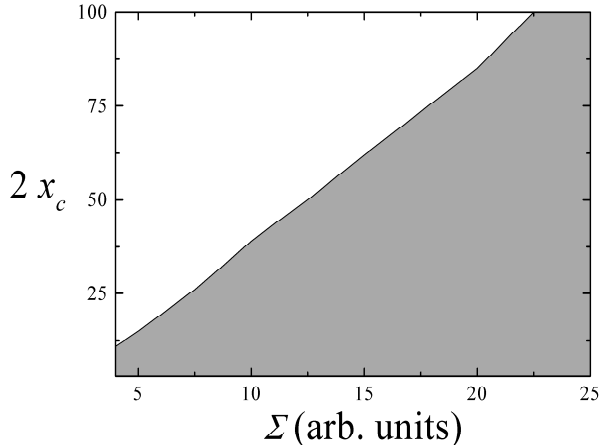}}}
\caption{Separation of large-amplitude patterns (shadowed region) from
negligible-amplitude patterns for the cut-off Gaussian influence functions
with width $\Sigma$ and cut-off length $x_c$. The domain size is  $L=100$.
Units for $\Sigma$ and $x_c$ are
arbitrary but identical. }\label{xsplane}
\end{figure}

In order to understand the interplay of $\Sigma$ and $x_c$ better, we vary
them independently and construct a phase plane $(\Sigma, x_c)$ in Fig.
\ref{xsplane}. The separation of the region of patterns with large amplitude
(displayed shadowed) from  that of patterns  of negligible amplitude is clear
and occurs along a line of slope $9/2$ \cite{foot3}.  This value of the slope
is significant in light of the fact that in Fig. 2, the transition is seen at
$\Sigma/L \approx 2/9.$ This clarifies that  it is the finite domain size that
imposes a cut-off on the Gaussians considered in Fig. 2, and that the domain
size $L$ there plays the precise role of $2x_c$ in Fig. 3.

The cut-off Gaussians or the square influence functions possess an abruptness
feature which would not be present in a physical system. To ensure smoothness,
we borrow from \cite{tsa} and consider an influence function (for periodic
boundary conditions)
\begin{eqnarray}
f_{\sigma ,\,r}(x)&=& \frac{1}{Z_{r}}[1-\frac{r\, x^2}{g(r,\sigma)}
]^{1/r} \Theta(\xi_c-x) \Theta(\xi_c+x),  \nonumber \\
 Z_{r}&=&\sqrt{\frac{\pi g(r,\sigma)}{r}}\frac{\Gamma
(1/r+1)}{\Gamma(1/r+3/2)}, \nonumber \\
 g(r,\sigma)&=&(2+3r) \sigma^2 \label{tsl}.
\end{eqnarray}
Here $r$ is non-negative, $\Gamma$ is the Gamma function, and
the influence function has a cut-off at $\xi_{c}$ with
\begin{equation}
\xi_{c}=\sigma \sqrt{\frac{2+3r}{r}}.
\end{equation}

\begin{figure}[hbt]
\centering \resizebox{\columnwidth}{!} {\rotatebox[origin=c]{-90}{
\includegraphics{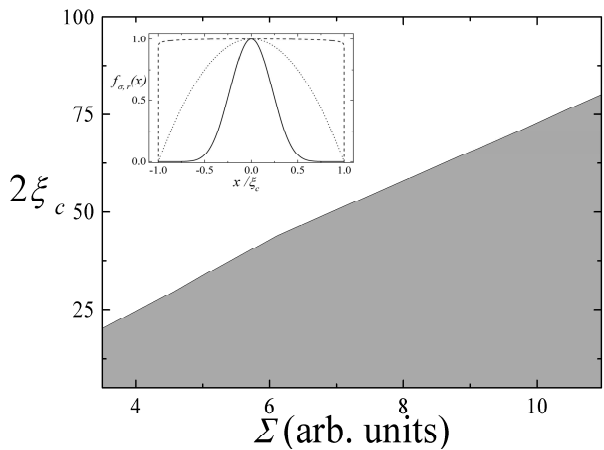}}}
\caption{Counterpart of Fig. 3 for the influence function of Eq. (\ref{tsl}).
 Inset shows $f_{\sigma ,\,r}$ in the periodic
boundary condition case for several values of $r=100$ (dashed), $1$ (dotted),
and $.01$ (full); respectively, $\Sigma=70$, $7$, and $.7$. The square and the
Gaussian emerge as particular cases. Units as in Fig. 3. } \label{xsplane2}
\end{figure}

The limit $r=0$ is the Gaussian. Note that $\xi_c$ denotes a \emph{natural}
cut-off of this function while $x_c$ in the Gaussian case is imposed externally by multiplying by a
symmetric square function of width 2$x_c$. The width $\Sigma$, the
range $\sigma$, and the parameter $r$ are related through
\begin{equation}
\Sigma=\xi_c\sigma(\xi_c^2-3\sigma^2)^{-1/2}=\xi_c\sqrt{2r}.
\end{equation}
 It is also possible to define the function $f_{\sigma ,\,r}$
for negative values of $r,$ specifically for $0>r>-2/3,$ through $f_{\sigma
,\,r}(x, y)= \frac{1}{Z_{r}}\left[
1-\frac{r(x-y)^2}{g(r,\sigma)}\right]^{1/r};\, Z_{r}=\sqrt{\frac{\pi
g(r,\sigma)}{-r}}\frac{\Gamma (-1/r-1/2)}{\Gamma(-1/r)} $ as in ref.
\cite{tsa}, but analysis with such long-tailed $f_{\sigma ,\,r}$'s does not
add anything important to our present study, the results being similar to
those for Gaussians. In addition to the smoothness property of these
functions, they have the feature that they reduce to a square or a Gaussian in
the respective limits $r\rightarrow 0$ and $r\rightarrow \infty$ (see inset in
Fig.  \ref{xsplane2}). It is therefore possible with their help to study for a
given cut-off length the effect of varying the width (central curvature) and
vice-versa. We fix a value of the natural cut-off $\xi_c$ well within the
domain size $L$, and plot in Fig. 4 the counterpart of Fig. 3 for the class of
influence functions given by Eq. (\ref{tsl}). The two plots are similar in
that the regions of large-amplitude and small-amplitude patterns are separated
cleanly by what appears to be a straight line in the phase space. The slopes
are different ($9/2$ for Fig. 3 but about $8$ for Fig. 4). We conclude that
the cut-off length and the central curvature are both relevant to the
formation of patterns in an interrelated manner shown by the phase plots. It
should be noticed that while the Gaussian has an infinite cut-off $x_c$ and a
finite width $\Sigma$,  the square has a finite cut-off $x_c$ and an infinite
width (curvature at center). Our introduction of Gaussians with an external
cut-off provided in the first part of our investigation Gaussians wherein both
control quantities were finite. Our introduction of the influence functions
defined in Eq. (\ref{tsl}) similarly provided square-like functions wherein
both control quantities were finite. The inset of Fig. 4 makes it particularly
clear that we can produce influence functions for this case with a fixed
cut-off and $\Sigma$ ranging from $0$ to $\infty$.

We have performed a number of further studies of pattern formation
from the long-range Fisher equation (2). They include extension to
\begin{figure}[hbt] \centering
\resizebox{\columnwidth}{!} {\rotatebox[origin=c]{-90}{
\includegraphics{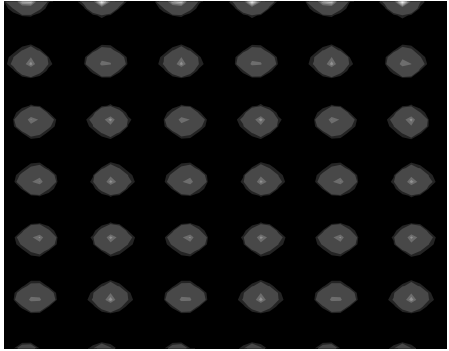}}}
\caption{Typical steady-state pattern in a 2D system with periodic boundary
conditions and a cut-off influence function. Lighter areas correspond to larger values of $u$.
Parameters are arbitrary.} \label{d2acotado}
\end{figure}
2-dimensional systems, typified by Fig. \ref{d2acotado}, in which patterns are
shown for periodic boundary conditions and a cut-off function of the kind
described in Eq. (\ref{tsl}). We have also studied the effects of boundary
conditions. As an example we display in Fig. \ref{cutoff1d} the lower-symmetry
\begin{figure}[hb]
\centering \resizebox{\columnwidth}{!}{\rotatebox[origin=c]{-90}{
\includegraphics{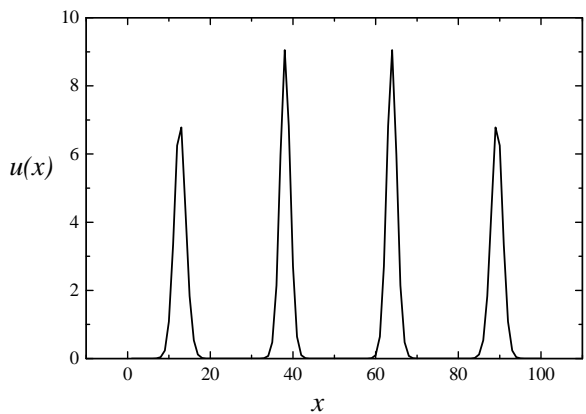}}}
\caption{Typical steady state patterns with cut-off influence function and
zero-flux boundary conditions.} \label{cutoff1d}
\end{figure}
patterns for zero-flux conditions that appear to have additional modulation in
the peaks. We have also investigated the time evolution of the patterns and
found it to have considerable complexity even for simple initial conditions.
We will report these various results elsewhere. Here we would like to draw
attention to the features of the influence function, cut-off length and
central curvature or width, whose ratio appears to determine whether the
steady-state patterns have negligible or sizeable amplitude, the critical value
of that ratio being different for different families of influence functions.
We have systematized our numerical findings through the phase plots in Figs.
\ref{xsplane} and \ref{xsplane2}. Our use of simple influence functions, such
as the square, the Gaussian, and the cut-off Gaussian, as well as of richer
functions as in ref. \cite{tsa} which reduce to these forms, has resulted in
our being able to focus on what features are responsible for what aspects of
the patterns. We hope that future investigations will clarify at a basic level
the picture presented by our numerical findings and make possible predictable
manipulation of patterns in real systems such as bacterial colonies by
controlling the influence function via control of the flow of nutrients and/or
chemotactic substances.

\section{Acknowledgements}

This work is supported in part by the Los Alamos National Laboratory via a
grant made to the University of New Mexico (Consortium of The Americas for
Interdisciplinary Science) and by National Science Foundation's Division of
Materials Research via grant No DMR0097204. 


\begin{thebibliography}{99}
\bibitem{foot} Such a critical range does not appear in the zero-flux case.
\end{thebibliography}
\end{document}